\def \AAP #1 #2 {{\em Astron. Astrophys.\/} {\bf #1}, #2}   
\def \AAL #1 #2 {{\em Astron. Astrophys. Lett.\/} {\bf #1}, L#2}   
\def \AAR #1 #2 {{\em Astron. Astrophys. Rev.\/} {\bf #1}, #2}   
\def \AAS #1 #2 {{\em Astron. Astrophys. Suppl. Ser.\/} {\bf #1}, #2}   
\def \AJ #1 #2 {{\em Astron. J.\/} {\bf #1}, #2}   
\def \ANNREV #1 #2 {{\em Ann. Rev. Astron. Astrophys.\/} {\bf #1}, #2}   
\def \APJ #1 #2 {{\em Astrophys. J.\/} {\bf #1}, #2}   
\def \APJL #1 #2 {{\em Astrophys. J. Lett.\/} {\bf #1}, L#2}   
\def \APJS #1 #2 {{\em Astrophys. J. Suppl.\/} {\bf #1}, #2}   
\def \APSS #1 #2 {{\em Astrophys. Space Sci.\/} {\bf #1}, #2}   
\def \ASR #1 #2 {{\em Adv. Space Res.\/} {\bf #1}, #2}   
\def \MN #1 #2 {{\em Mon. Not. R. Astr. Soc.\/} {\bf #1}, #2}   
\def \PRL #1 #2 {{\em Phys. Rev. Lett.\/} {\bf #1}, #2}   
\def \NAT #1 #2 {{\em Nature\/} {\bf #1}, #2}   

\documentclass[12pt, a4paper]{article} 
\input epsf.sty   
 
\begin{document}   
 
\title{Cosmic Ray Acceleration at Relativistic Shocks} 
   
\author{Micha{\l} Ostrowski \\ 
Obserwatorium Astronomiczne, Uniwersytet Jagiello\'nski, \\ 
ul. Orla 171, 30-244 Krak\'ow, Poland  (E-mail:  mio{@}oa.uj.edu.pl)}   
   
\maketitle

\begin{abstract}   
Theoretical studies of cosmic ray particle acceleration in the
first-order Fermi process at relativistic shocks are reviewed. At the
beginning we discuss the acceleration processes acting at {\it mildly}
relativistic shock waves.  An essential role of oblique field
configurations and field perturbations in forming the particle energy
spectrum and changing the acceleration time scale is discussed. Then, we
report on attempts to consider particle acceleration at {\it
ultra-relativistic} shocks, often yielding an asymptotic spectral index
$\sigma \approx 2.2$ at large shock Lorentz factors. We explain why this
result is limited to the cases of highly turbulent conditions near
shocks. We conclude that our present knowledge of the acceleration
processes acting at relativistic shocks is insufficient to allow for
realistic modelling of the real shocks. The present review is a
modified, extended and updated version of Ostrowski (1999).

\vspace{2mm}

\vspace{2mm}
\end{abstract}

\section{Introduction}   
   
Relativistic plasma flows are detected or postulated to exist in a
number of astrophysical objects, ranging from a mildly relativistic jet
of SS433, through the-Lorentz-factor-of-a-few jets in AGNs and galactic
`mini-quasars', up to ultra-relativistic outflows in sources of gamma
ray bursts and, possibly, in pulsar winds. As nearly all such objects
are efficient emitters of synchrotron radiation and/or high energy
photons requiring the existence of energetic particles, our attempts to
understand the processes generating cosmic ray particles are essential
for understanding the fascinating phenomena observed. Below we will
discuss the work carried out  in order to understand the cosmic ray
first-order Fermi acceleration processes acting at relativistic shocks.
One should note that in the present discussion we consider the high
energy particles with gyroradii (or mean free paths) much larger than
the shock thickness defined by the compressed `thermal' plasma. The
present review is an updated version of Ostrowski  (1999), also
including an extended discussion of the acceleration  processes acting
at ultra-relativistic shocks (Ostrowski \& Bednarz 2002).
   
\section{Particle acceleration at non-relativistic shock waves}   
   
Processes of the first-order particle acceleration at non-relativistic
shock waves were widely discussed by a number of authors during the last
two decades (for review, see, e.g. Drury 1983, Blandford \& Eichler
1987, Berezhko et al. 1988, Jones \& Ellison 1991). Below, we review
the basic physical picture and some important results obtained within
this theory {\it for test particles}, to be later compared with the
results obtained for relativistic shocks.
   
The simple description of the acceleration process preferred by us
consists of  considering two plasma rest frames, the {\it upstream
frame} and the {\it downstream one}. We use indices `$1$' or `$2$' to
indicate quantities measured in the upstream or the downstream frame
respectively. If one neglects the second-order Fermi acceleration, the
particle energy is a constant of motion in any of these plasma rest
frames and energy changes occur when the particle momentum is
Lorentz-transformed at each crossing of the shock. In the case of {\it
parallel} shock, with the mean magnetic field parallel to the shock
normal, the acceleration of an individual particle is due to the
consecutive shock crossings by the diffusive wandering particle. Each
{\it upstream-downstream-upstream} diffusive loop results in a small
increment of particle momentum, $\Delta p \propto p \cdot (U_1-U_2)/v$,
where $v$ is the particle velocity and $U_i$ is the shock velocity in
the respective $i = 1$ or $2$ frame, $U_1 \ll v $. One should note that
in oblique shocks, the particle helical trajectory can cross the shock
surface a number of times at any individual shock transition or
reflection.
   
The most interesting feature of the first-order Fermi acceleration at a
non-relativistic plane-parallel shock wave is the independence of the
{\it test-particle stationary} particle energy spectrum from the
background conditions near the shock, including the mean magnetic field
configuration and the spectrum of MHD turbulence. The main reason behind
that is a nearly-isotropic form of the particle momentum distribution at
the shock. If a sufficient amount of scattering occurs near the shock,
this condition always holds for the shock velocity along the upstream
magnetic field $U_{B,1} \equiv U_1 / \cos \Psi_1 \ll v$ ($\Psi_1$ - the
upstream magnetic field inclination to the shock normal). Independently
of the field inclination at the shock, the particle density is
continuous across it and the spectral index for the phase-space
distribution function, $\alpha$, is given exclusively in the terms of a
single parameter -- the shock compression ratio $R$:
   
$$\alpha = {3R \over R-1}  \qquad   . \eqno(2.1) $$   
   
\noindent   
Because of the isotropic form of the particle distribution function, the
spatial diffusion equation has become a widely used mathematical tool
for describing particle transport and acceleration processes in
non-relativistic flows. With its use the characteristic acceleration
time scale at the parallel ($\Psi_1=0$) shock can be derived as
   
$$T_{acc} = {3 \over U_1-U_2}\, \left\{ {\kappa_1 \over U_1} + {\kappa_2   
\over U_2} \right\} \qquad , \eqno(2.2) $$   
   
\noindent   
where $\kappa_i \equiv \kappa_{\parallel ,i}$ is the respective particle
spatial diffusion coefficient along the magnetic field, as discussed by
e.g. Lagage \& Cesarsky (1983). Ostrowski (1988a; see also Bednarz \&
Ostrowski 1996) derived an analogous expression for shocks with oblique
magnetic fields and small amplitude magnetic field perturbations. For a
negligible cross-field diffusion and for $U_{B,1} \ll c$ it can be
written in essentially the same form as the one given in Eq.~(2.2), with
all quantities taken as the normal ($n$) ones with respect to the shock
($\kappa_{n,i}$ for $\kappa_i$ ($i$ = $1$, $2$)). As $\kappa_n <
\kappa_\parallel$, the oblique shocks may be more rapid accelerators
when compared to the parallel shocks.
   
Not discussed here non-linear and time dependent effects, inclusion of
additional energy losses and gains, etc., make the physics of the
acceleration more intricate, allowing e.g. for non-power-low and/or
non-stationary particle distributions.
   
\section{Cosmic ray acceleration at relativistic shock waves}   
   
\subsection{The Fokker-Planck description of the acceleration process}   
   
In the case of the shock velocity (or its projection $U_{B,1}$) reaching
values comparable to the light velocity, the particle distribution at
the shock becomes anisotropic. This fact complicates to a great extent
both the physical picture and the mathematical description of particle
acceleration. The first attempt to consider the acceleration process at
the relativistic shock was presented in 1981 by Peacock (see also Webb
1985); however, no consistent theory was proposed until a paper of Kirk
\& Schneider (1987a; see also Kirk 1988) appeared. Those authors
considered the stationary solutions of the relativistic Fokker-Planck
equation for particle pitch-angle diffusion for the case of the parallel
shock wave. In the situation with the gyro-phase averaged distribution
$f(p, \mu, z)$, which depends only on the unique spatial co-ordinate $z$
along the shock velocity, and with $\mu$ being the pitch-angle cosine,
the equation takes the form:
   
$$\Gamma ( U + v \mu ) {\partial f \over \partial z} = C(f) + S \qquad ,   
 \eqno(3.1)  $$   
   
\noindent   
where $\Gamma \equiv 1/\sqrt{1-U^2}$ is the flow Lorentz factor, $C(f)$
is the collision operator and $S$ is the source function. In the
presented approach, the spatial co-ordinates are measured in the shock
rest frame, while the particle momentum co-ordinates and the collision
operator are given in the respective plasma rest frame. For the applied
pitch-angle diffusion operator, $C = \partial / \partial \mu (D_{\mu
\mu} \partial f / \partial \mu)$, they generalised the diffusive
approach to higher order terms in particle distribution anisotropy and
constructed general solutions at both sides of the shock which involved
solutions of the eigenvalue problem. By matching two solutions at the
shock, the spectral index of the resulting power-law particle
distribution can be found by taking into account a sufficiently large
number of eigenfunctions. The same procedure yields the particle angular
distribution and the spatial density distribution. The low-order
truncation in this approach corresponds to the standard diffusion
approximation and to a somewhat more general method described by
Peacock. The above analytic approach (or the `semi-analytic' one, as the
mentioned matching of two series involves numerical fitting of the
respective coefficients) was verified by Kirk \& Schneider (1987b) by
the method of particle Monte Carlo simulations.

\begin{figure}[hbt]   
\vspace{70mm} 
\includegraphics{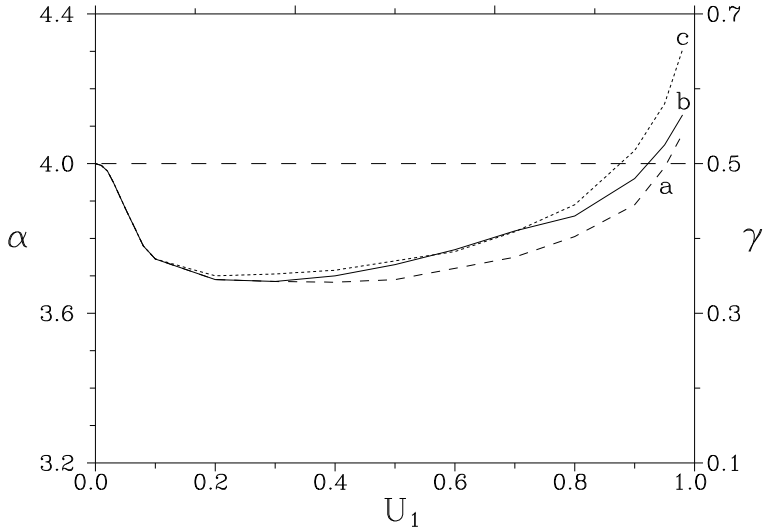}    
\vspace{0mm} 
\caption{  
The particle spectral indices $\alpha$ at parallel shock   
waves propagating in the cold ($e$, $p$) plasma versus the shock   
velocity $U_1$ (Heavens \& Drury 1988). On the right vertical axis the   
respective synchrotron spectral index $\gamma$ is given. Using the solid   
line (b) and the dashed line (a) we show indices for two choices of the   
turbulence spectrum. The dashed line (c) gives the spectral index   
derived from Eq.~2.1. The horizontal line $\alpha = 4.0$ is given for   
the reference.} \end{figure}   
   
An application of this approach to more realistic conditions -- but
still for parallel shocks -- was presented by Heavens \& Drury (1988),
who investigated the fluid dynamics of relativistic shocks (cf. also
Ellison \& Reynolds 1991) and used the results to calculate spectral
indices for accelerated particles (Fig.~1). They considered the shock
wave propagating into electron-proton or electron-positron plasma, and
performed calculations using the analytic method of Kirk \& Schneider
for two different power spectra for the scattering MHD waves. In
contrast to the non-relativistic case, they found (see also Kirk 1988)
that the particle spectral index depends on the form of the wave
spectrum.  The unexpected fact was noted that the non-relativistic
expression (2.1) provided a quite reasonable approximation to the actual
spectral index.
   
\begin{figure}[hbt]   
\vspace{70mm} 
\includegraphics{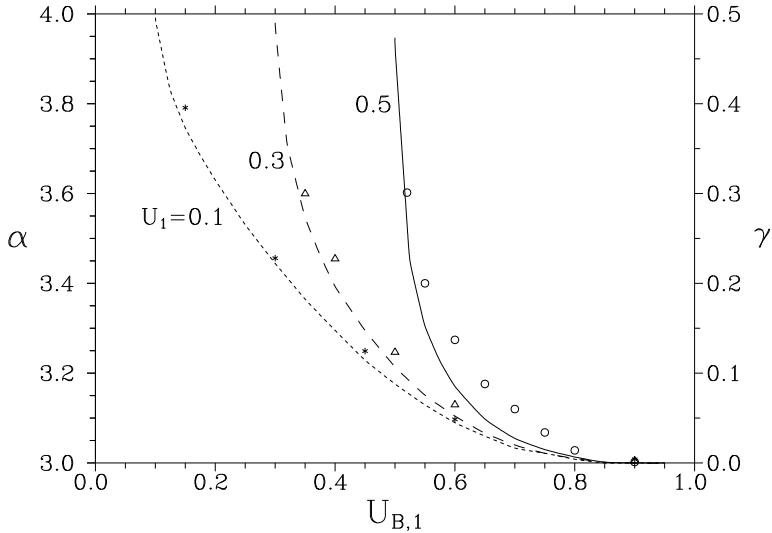}    
\vspace{0mm} 
\caption{  
Spectral indices $\alpha$ of particles accelerated at oblique shocks   
versus shock velocity projected at the mean magnetic field, $U_{B,1}$.   
On the right the respective synchrotron spectral index $\gamma$ is   
given. The shock velocities $U_1$ are given near the respective curves   
taken from Kirk \& Heavens (1989). The points were taken from   
simulations deriving explicitly the details of particle-shock   
interactions (Ostrowski 1991a). The results are presented for   
compression $R = 4$.} \end{figure}   
   
A substantial progress in understanding the acceleration process in the
presence of highly anisotropic particle distributions is due to the work
of Kirk \& Heavens (1989; see also Ostrowski 1991a and Ballard \&
Heavens 1991), who considered particle acceleration at {\it subluminal}
($U_{B,1} < c$) relativistic shocks with oblique magnetic fields. They
assumed the magnetic momentum conservation, $p_\perp^2/B = const$, at
particle interaction with the shock and applied the Fokker-Planck
equation discussed above to describe particle transport along the field
lines outside the shock, while excluding  the possibility of cross-field
diffusion. In the cases when $U_{B,1}$ reached relativistic values, they
derived very flat energy spectra with $\gamma \approx 0$ at $U_{B,1}
\approx 1$ (Fig.~2). In such conditions, the particle density in front
of the shock can substantially -- even by a few orders of magnitude --
exceed the downstream density (see the curve denoted `-8.9' at Fig.~3).
Creating flat spectra and great density contrasts is due to the
effective reflections of anisotropically distributed upstream particles
from the region of compressed magnetic field downstream of the shock.
However, the conditions leading to very flat spectra are supposed to be
accompanied by processes -- like a large amplitude wave generation
upstream of the shock -- leading to spectrum steepening (cf. Sec.~3.2).
   
As stressed by Begelman \& Kirk (1990), in relativistic shocks one can
often find the {\it superluminal} conditions with $U_{B,1} > c$, where
the above presented approach is no longer valid. Then, it is not
possible to reflect upstream particles from the shock and to transmit
downstream particles into the upstream region. In effect, only a single
transmission of upstream particles re-shapes the original distribution
by shifting particle energies to larger values. The energy gains in such
a process, involving a highly anisotropic particle distribution, can be
quite significant, exceeding the value expected for the adiabatic
compression.
   
The approach proposed by Kirk \& Schneider (1987a) and Kirk \& Heavens
(1989), and the derivations of Begelman \& Kirk (1990) are valid only in
case of weakly perturbed magnetic fields. However, in the efficiently
accelerating shocks one may expect  large amplitude waves to be present,
when both the Fokker-Planck approach is no longer valid and the magnetic
momentum conservation no longer holds for oblique shocks. In such a
case, numerical methods have to be used.
   
\subsection{Particle acceleration in the presence of large amplitude   
magnetic field perturbations}   
   
\begin{figure}[hbt]   
\vspace{76mm} 
\includegraphics{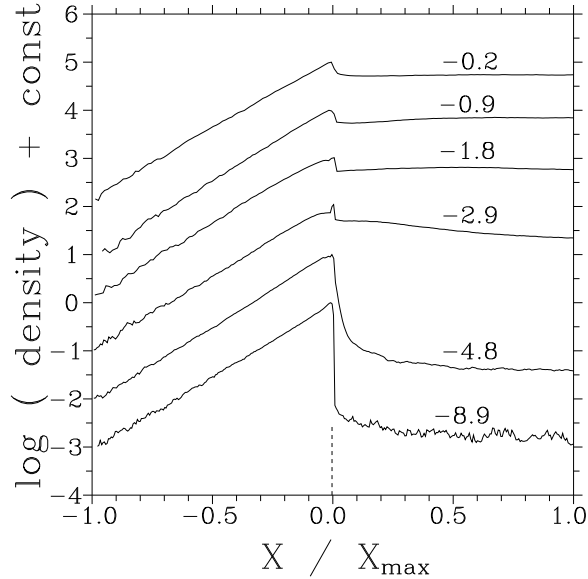}    
\vspace{0mm} 
\caption{  
The energetic particle density across the relativistic shock   
with an oblique magnetic field (Ostrowski 1991b). The shock with $U_1 =   
0.5$, $R = 5.11$ and $\psi_1 = 55^o$ is considered. The curves for   
different perturbation amplitudes are characterized with the value $\log   
\kappa_\perp / \kappa_\parallel$ given near the curve. The data are   
vertically shifted for picture clarity. The value $X_{max}$ is the   
distance from the shock at which the upstream particle density decreases   
to $10^{-3}$ part of the shock value.} \end{figure}   
   
The first attempt to consider the acceleration process at parallel shock
wave propagating in a turbulent medium was presented by Kirk \&
Schneider (1988), who included into Eq.~3.1 the Boltzmann collision
operator describing the large angle scattering. By solving the resulting
integro-differential equation they demonstrated the hardening of the
particle spectrum due to increasing contribution of the large-angle
scattering. The reason for such a spectral change is the additional
isotropization of particles interacting with the shock, leading to an
increase in the particle mean energy gain. In oblique shocks, this
simplified approach cannot be used because the character of individual
particle-shock interaction -- reflection and transmission
characteristics -- depends on the magnetic field perturbations. Let us
additionally note that application of the point-like large-angle
scattering model in relativistic shocks does not provide a viable
physical representation of the scattering at MHD waves (Bednarz \&
Ostrowski 1996).
   
\begin{figure}[hbt]   
\vspace{71mm} 
\includegraphics{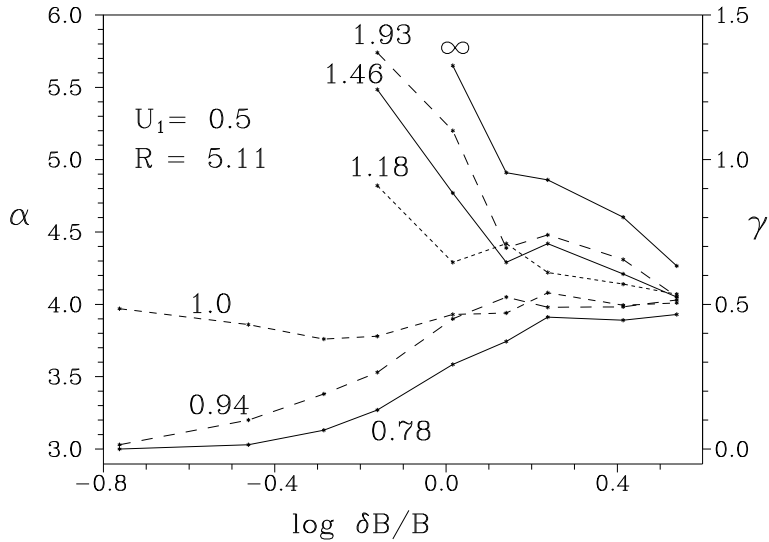}    
\vspace{0mm} 
\caption{  
Spectral indices for oblique relativistic shocks versus   
perturbation amplitude $\delta B/B$ (Ostrowski 1993). Different field   
inclinations are characterized by the values of $U_{B,1}$ given near the   
respective results, $U_{B,1} < 1$ for subluminal shocks and $U_{B,1} \ge   
1$ for superluminal ones. Absence of data for small field amplitudes in   
superluminal shocks is due to extremely steep power law spectra   
occurring in these conditions (cf. Begelman \&  Kirk 1990). Decreasing   
the field inclination $\Psi_1 \to 0$ (i.e. to the parallel shock with   
$U_{B,1} = U_1$) gives spectral indices more and more similar to a   
constant line $\alpha = 3.72$,  not shown here   
for picture clarity (cf. Fig-s~1,2).} \end{figure}   
   
To handle the problem of the particle spectrum in a wide range of
background conditions, the Monte Carlo particle simulations were
proposed (Kirk \& Schneider 1987b; Ellison et al. 1990; Ostrowski 1991a,
1993; Ballard \& Heavens 1992, Naito \& Takahara 1995, Bednarz \&
Ostrowski 1996, 1998). At first, let us consider subluminal shocks. The
field perturbations influence the acceleration process in various ways.
As they enable the particle cross field diffusion, a modification
(decrease) of the downstream particle's escape probability may occur.
This factor tends to harden the spectrum. Next, the perturbations
decrease particle anisotropy, leading to an increase of the mean energy
gain of reflected upstream particles, but -- what is more important for
oblique shocks -- this also increases the particle upstream-downstream
transmission probability due to less efficient reflections, enabling
them to escape from further acceleration. The third factor is due to
perturbing particle trajectory during an individual interaction with the
shock discontinuity and breakdown of the approximate conservation of
$p_\perp^2/B$. Because reflecting a particle from the shock requires a
fine tuning of the particle trajectory with respect to the shock
surface, even small amplitude perturbations can decrease the reflection
probability in a substantial way. Simulations show (see Fig.~4 for
$U_{B,1} < 1.0$) that -- until the wave amplitude becomes very large --
the factors leading to efficient particle escape dominate with the
resulting steepening of the spectrum to $\gamma \sim 0.5$ -- $0.8$, and
the increased downstream transmission probability lowers the cosmic ray
density contrast across the shock (Fig.~3).
   
In parallel shock waves propagating in a highly turbulent medium, the
effects discovered for oblique shocks can also manifest their  presence
because of the {\it local} perturbed magnetic field compression at the
shock. The problem was considered using the technique of particle
simulations by Ballard \& Heavens (1992; cf. Ostrowski 1988b for
non-relativistic shock). They showed a possibility of having a very
steep spectrum in this case, with the spectral index growing from
$\gamma \sim 0.6$ at medium relativistic velocities up to nearly $2.0$
at $U_1 = 0.98$. These results apparently do not correspond to the
large-perturbation-amplitude limit of Ostrowski's (1993; see the
discussion therein) simulations for oblique shocks and the analytic
results of Heavens \& Drury (1988).
   
For large amplitude magnetic field perturbations the acceleration
process in superluminal shocks can lead to the power-law particle
spectrum formation, against the statements of Begelman \& Kirk (1990)
valid at small wave amplitudes only. Such  a general case was discussed
by Ostrowski (1993; see Fig.~4 for $U_{B,1} \ge 1$) and by Bednarz \&
Ostrowski (1996, 1998).
   
\subsection{The acceleration time scale}   
   
\begin{figure}[hbt]   
\vspace{72mm} 
\includegraphics{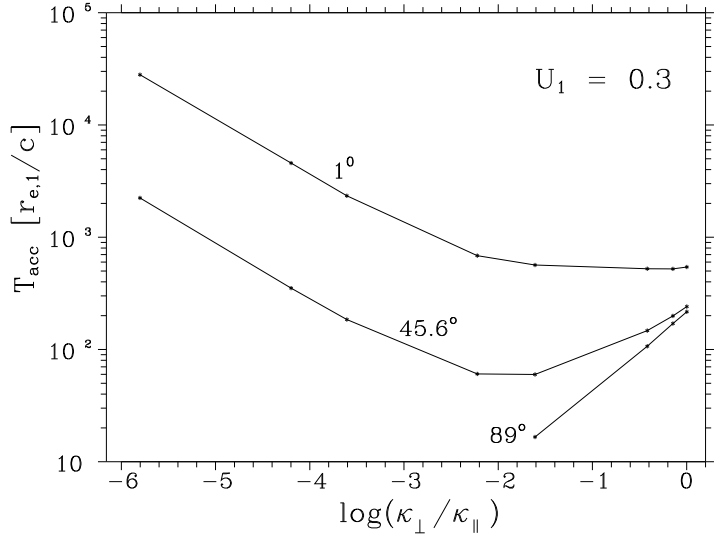}    
\vspace{0mm} 
\caption{  
The acceleration time $T_{acc}$ versus the level of particle   
scattering measured by the ratio of $\kappa_\perp / \kappa_\parallel$
(Bednarz \& Ostrowski 1996). We present results for three
values of the magnetic field inclination: a.) parallel shock ($\psi_1 =   
1^\circ$), b.) a subluminal shock with $\psi_1 = 45.6^\circ $ and c.) a   
superluminal shock with $\psi_1 = 89^\circ $. $r_{e,1}$ is the particle    
gyroradius in the effective (including   
perturbations) upstream magnetic field .}   
\end{figure}   
   
\noindent   
The shock waves propagating with relativistic velocities also raise
interesting questions pertaining to the cosmic ray acceleration time
scale, $T_{acc}$. A simple comparison to  non-relativistic values shows
that $T_{acc}$ relatively decreases with increasing shock velocity for
parallel (Quenby \& Lieu 1989; Ellison et al. 1990) and oblique
(Takahara \& Terasawa 1990; Newman et al. 1992; Lieu et al. 1994; Quenby
\& Drolias 1995; Naito \& Takahara 1995) shocks. However, the numerical
approaches used there, based on assuming particle isotropization for all
scatterings, neglect or underestimate a significant factor affecting the
acceleration process -- the particle anisotropy. Ellison et al. (1990)
and Naito \& Takahara (1995) also included the more realistic, in our
opinion, derivations involving the pitch-angle diffusion approach. The
calculations of Ellison et al. for parallel shocks show similar results
to those they obtained for large amplitude scattering. For the shock
with velocity $0.98\,c$ the acceleration time scale is reduced by the
factor $\sim 3$ with respect to the non-relativistic formula of
Eq.~2.2~. Naito \& Takahara considered shocks with oblique magnetic
fields. They confirmed the reduction of the acceleration time scale with
an increasing inclination of the magnetic field, derived earlier for
non-relativistic shocks. However, their approach neglected effects of
particle cross field diffusion and assumed the adiabatic invariant
conservation in particle interactions with the shock, thus limiting the
validity of their results to a small amplitude turbulence near the
shock.
   
\begin{figure}[hbt]   
\vspace{72mm} 
\includegraphics{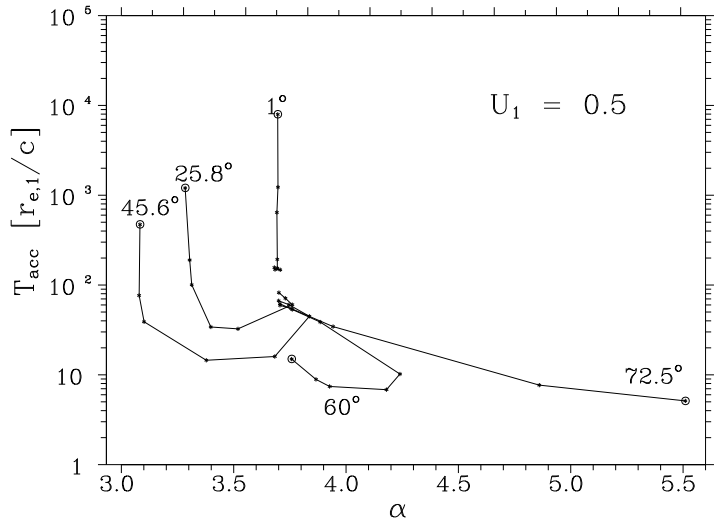}    
\vspace{0mm} 
\caption{  
The relation of $T_{acc}$ versus the particle spectral index   
$\alpha$ at different magnetic field inclinations $\psi_1$ given near   
the respective curves. The {\it minimum} value of the model parameter   
$\kappa_\perp/\kappa_\|$ occurs at the encircled point of each curve and   
the wave amplitude monotonously increases along each curve up to $\delta   
B \sim B$; $r_{e,1}$ -- see Fig.~5.} \end{figure}   
   
A wider discussion of the acceleration time scale is presented by
Bednarz \& Ostrowski (1996), who apply numerical simulations involving
the small angle particle momentum scattering. The approach is also
believed to provide a reasonable description of particle transport in
the presence of large $\delta B$, and thus to enable modelling of the
effects of cross-field diffusion. The resulting values (Fig-s~5, 6) are
given in the shock {\it normal} rest frame (cf. Begelman \&  Kirk 1990).
In parallel ($\Psi_1 = 1^\circ$) shocks $T_{acc}$ diminishes with the
growing perturbation amplitude and  shock velocity $U_1$. However, it is
approximately constant for a given value of $U_1$ if we use the formal
diffusive time scale, $\kappa_1/(U_1c) + \kappa_2/(U_2c)$, as the time
unit. A new feature discovered in oblique shocks is that due to the
cross-field diffusion $T_{acc}$ can change with $\delta B$ in a
non-monotonic way (Fig.~5). The acceleration process leading to the
power-law spectrum is possible in superluminal shocks only in the
presence of large amplitude turbulence. Then, in contrast to the
quasi-parallel shocks, $T_{acc}$ increases with increasing $\delta B$.
In the considered cases with the oblique field configurations one may
note a possibility to have an extremely short acceleration time scale
comparable to the particle gyroperiod in the magnetic field upstream of
the shock. A coupling between the acceleration time scale and the
particle spectral index is presented in Fig.~6. One should note that the
form of involved relation is contingent to a great extent on the
magnetic field configuration.

\section{Energy spectra of cosmic rays accelerated at large   
Lorentz-factor shocks} 
 
Ultra-relativistic shock waves suggested to be sources of gamma-ray
bursts are also expected by some authors to produce ultra-high-energy
cosmic ray particles. The process of the first-order Fermi acceleration
in such shocks was discussed in a series of papers by Bednarz \&
Ostrowski (1997, 1998; see also Bednarz 2000a,b), Gallant \& Achterberg
(1999; see also Achterberg et al. 2001), Kirk et al. (2000) and Vietri
(2000). Below, following  Ostrowski \& Bednarz (2002) we shortly compare
and discuss different approaches to the considered acceleration process.

\subsection{The first-order Fermi acceleration at ultra-relati\-vistic 
shocks} 
 
The first-order Fermi acceleration process at an ultra-relativistic
shock wave involves extreme particle anisotropy at the shock in the
upstream plasma rest frame (UPF), and more mild distributions in the
shock normal rest frame or the downstream plasma rest frame (cf.
Begelman \& Kirk 1990). Let us consider an individual cosmic ray
particle acceleration starting with a particle crossing the shock
upstream (cf. Gallant \& Achterberg 1999). Then, in UPF, its momentum is
nearly parallel to the shock normal. When the shock Lorentz factor is
large ($\Gamma \gg 1$) the particle stays in front of the shock for a
time required for a slight, $\sim 1/\Gamma$, deflection of its momentum
allowing the shock to overtake it and transmit to the downstream region.
The deflection proceeds due to the magnetic field upstream of the shock,
consisting of the large scale smooth background structure perturbed by
the MHD fluctuations. This tiny change of particle momentum upstream of
the shock allows for its transmission downstream of the shock, where --
due to the Lorentz transformation with a large $\Gamma$ -- its momentum
direction can be changed at a large angle with respect to its original
direction before the transmission upstream. Such large amplitude angular
scatterings can enable a finite fraction of particles to follow
trajectories leading to successive transmissions upstream of the
shock. Repeating of the described loops, with each roughly doubling the
particle energy, leads to formation of the power law particle spectrum.
Several authors (Bednarz \& Ostrowski 1998, Gallant \& Achterberg 1999,
Gallant et al. 1999) discussed this process leading to formation of the
spectrum with the energy spectral index $\sigma \approx 2.2$ at $\Gamma
\gg 1$~. Essentially the same results were obtained within different
approaches presented by the above authors and by Kirk et al. (2000) and
Vietri (2000).

The work of Bednarz \& Ostrowski (1997, 1998) was based on Monte Carlo
simulations of particle transport governed by small amplitude pitch
angle scattering. Thus, depending on the  mean time between successive
scattering acts, $\Delta t$, and  the maximum angular scattering
amplitude, $\Delta \Omega_{max}$, it was possible to model situations
with different mean field configurations and different amounts of
turbulence. The mean field configuration downstream of the shock was
derived from the mean upstream field using the appropriate jump
conditions and trajectories of particles interacting with the shock
discontinuity were derived exactly for such fields. The approach takes
into account correlations in the process due to the regular part of the
magnetic field, but irregularities responsible for pitch angle
scattering are introduced as random. In order to model particle pitch
angle diffusion upstream of the shock, with nearly a delta-like angular
distribution  an extremely small scattering amplitude should be used,
$\Delta \Omega_{max} \ll \Gamma^{-1}$.  Increasing the shock Lorentz
factor results in decreasing the momentum perturbation required for its
transmission downstream and leaves a shorter time for this perturbation,
$t_1$. In the applied pitch angle diffusion approach the momentum
variation due to the regular component of the magnetic field scales like
$t_1$, whence the diffusive change scales like $t_1^{1/2}$. Thus growing
$\Gamma$ leads to decreasing $t_1$ and the diffusive term have to
dominate at sufficiently large $\Gamma$. However, one should note that
with decreasing $\Delta t$ and $\Delta \Omega_{max}$, when the
interaction proceeds at the sub-resonance ($\ll r_g$) spatial scale, a
serious physical problem with the applied approach appears: it requires
the large amplitude short wave turbulence to be non-linear at shortest
scales.

An analogous, pitch angle diffusion modelling appended considerations of
Gallant et al. (1999; for a more detailed description see Achterberg et
al. 2001), who obtained essentially the same spectral indices as the
asymptotic one derived by Bednarz \& Ostrowski (1998). They considered a
highly turbulent conditions near the shock leading to the particle pitch
angle diffusion {\it with respect to the shock normal}, i.e. the regular
part of the magnetic field (or continuity of the field across the shock)
was neglected.  Thus, for example, if the amplitude of the magnetic
field turbulence is limited, it can not reproduce spectrum steepening
(or flattening at intermediate Lorentz factors) in the presence of
oblique magnetic fields (cf. Ostrowski 1993, Bednarz \& Ostrowski 1998,
Begelman \& Kirk 1990). The both above models describe essentially the
same physical situation only for shocks propagating in the highly
turbulent medium.

An alternative discussion of the acceleration process presented by
Gallant \& Achterberg (1999) was based on a simple turbulence model. In
their approach a highly turbulent magnetic field configuration was
assumed upstream and downstream of the shock, idealized as cells filled
with randomly oriented uniform magnetic fields. With such approach
particles crossing the shock enter a new cell with a randomly selected
magnetic field configuration. Thus, there always occur configurations
allowing some particles crossing the shock downstream to reach it again
and to form the power law spectrum. In this model there is no need for
the upstream magnetic field perturbations and a model with the uniform
upstream field yields the same power law distribution.

Two quasi-analytic approaches to the considered acceleration process
were presented by Kirk et al. (2000) and Vietri (2000). Both attempt to
solve the Fokker-Planck equation describing particle advection with the
general plasma flow and the small amplitude scattering of particle pitch
angle as measured with respect {\it to the shock normal}. The important
work of Kirk et al. modified the Kirk \& Schneider (1987a) series
expansion approach to treat the delta-like angular distribution upstream
of the shock. An analytically more simple Vietri approach applies
convenient ansatz'es for the anisotropic upstream and downstream
particle distributions, resembling the Peacock's (1981) approach to
acceleration at `ordinary' relativistic shocks. Both methods confirm the
results of the earlier numerical modelling. A deficiency of the above
semi-analytic approaches is its' inability to treat situations with
mildly perturbed magnetic fields, on average oblique to the shock
normal. If considered valid for different magnetic field configurations
these models require the large amplitude short wave turbulence to remove
any signature of the uniform background field or of the long wave
perturbations.

\subsection{Acceleration at the ultra-relativistic shock near the   
Crab Pulsar}   
   
In the discussion above, to treat the shock as the flow discontinuity,
and the acceleration process to be of the first-order Fermi type, one
had to consider very high energy particles. Quite interesting
alternative approach  intended to study the acceleration process
starting from low `thermal' energies was proposed by Hoshino et al.
(1992; see also Gallant \& Arons 1994; for review Arons 1996). They
considered acceleration at the ultra-relativistic shock formed in the
wind outflow of the (e$^+$,e$^-$) pair plasma containing heavy nuclei
and being permeated by the weak magnetic field oriented perpendicular to
the flow direction, i.e. in a model wind for the Crab Pulsar. In the
large Lorentz factor wind, the ram pressure of nuclei dominates over the
ram pressure of the pair plasma, and both these pressures are much
larger than the magnetic field pressure.
   
At the collisionless shock, the pairs' bulk velocity is isotropiz\-ed
much more efficiently, leaving nuclei penetrating the downstream region
as a particle beam. This process generates an electric field in the
shock and -- due to the ion distribution anisotropy -- generates long
electromagnetic plasma waves. Damping of such waves by pairs accelerates
some of electrons/positrons to energies comparable to the iron nuclei
energies downstream of the shock. The work mentioned here is based on
the results of numerical plasma  simulations of the ultra-relativistic
collisionless shock.

\section{Final remarks}   
   
One may note that observations of possible sites of relativistic shock
waves (knots and hot spots in extragalactic radio sources), which allow
for the determination of the energetic electron spectra, often yield
particle spectral indices close to $\alpha = 4.0$ ($\gamma = 0.5$). The
theoretical work done to date on the {\it test particle} cosmic ray
acceleration at mildly relativistic shocks yield not too promising
results for meaningful modelling of these astrophysical sources. The main
reason for this deficiency is -- in contrast to the non-relativistic
shocks -- a direct dependence of the derived spectra on the conditions
near the shock. Not only the shock compression ratio, but also other
parameters, like the mean inclination of the magnetic field or the
turbulence spectrum and its amplitude, are significant. Depending on the
actual conditions one may obtain spectral indices as flat as $\alpha =
3.0$ ($\gamma = 0.0$) or very steep ones with $\alpha > 5.0$ ($\gamma >
1.0$). The background conditions leading to the very flat spectra are
probably subject to some instabilities; however, there is no detailed
derivation describing the instability growth and the resulting cosmic
ray spectrum modification.
  
The situation was supposed to be simpler for  large $\Gamma$ shocks,
where the spectral index seems to converge to the universal limit
$\sigma_\infty \approx 2.2$.   However, as pointed out above, the
validity of this result may be quite limited. In this moment it is
difficult to evaluate if the required conditions are satisfied at
the studied ultra-relativistic shocks.
   
A true progress in modelling particle acceleration in actual sources
requires a full plasma non-linear description (see also Ostrowski 1994),
including the second-order acceleration processes and a feedback of
accelerated particles at the turbulent wave fields near the shock wave,
the flow modification caused by the cosmic rays' plasma pre-shock
compression and, of course, the appropriate boundary conditions. A
simple non-linear approach to the parallel shock case was presented by
Baring \& Kirk (1990), who found that relativistic shocks could be very
efficient accelerators. However, it seems to us that in a more general
case it will be very difficult to make any substantial progress in that
matter. For very flat particle spectra the non-linear acceleration
picture depends to a large extent on the detailed knowledge of the
background and boundary conditions in the scales relevant for particles
near the upper energy cut-off. The existence of stationary solutions is
doubtful in this case. A noticeable progress in considering detailed
physics of the acceleration in relativistic collisionless shocks may
result from application of the particle-in-cell simulations, including
physical processes (instabilities) discussed by e.g. Hoshino et al.
(1992), Medvedev \& Loeb (1999) or Pohl et al. (2002).

The present work was supported by the {\it Komitet Bada\'n Naukowych}
through the grant PB 258/P03/99/17.

\end{document}